\documentclass[aps,prl,twocolumn,
comment, 
superscriptaddress,groupedaddress,
amsmath,amssymb,floatfix,footinbib]
{revtex4}  
\usepackage{float}
\usepackage{graphicx}
\usepackage{bm}% bold math
\usepackage[usenames,dvipsnames]{color}
\usepackage{slashed}
\usepackage{slashed}

\begin{document}

\renewcommand{\(}{\left(}
\renewcommand{\)}{\right)}
\renewcommand{\{}{\left\lbrace}
\renewcommand{\}}{\right\rbrace}
\renewcommand{\[}{\left\lbrack}
\renewcommand{\]}{\right\rbrack}
\renewcommand{\Re}[1]{\mathrm{Re}\!\{#1\}}
\renewcommand{\Im}[1]{\mathrm{Im}\!\{#1\}}
\newcommand{\dd}[1][{}]{\mathrm{d}^{#1}\!\!\;}
\newcommand{\del}{\partial}
\newcommand{\nn}{\nonumber}
\newcommand{\ie}{i.e.\,}
\newcommand{\cf}{cf.\,}
\newcommand{\refeq}[1]{Eq.~(\ref{eq:#1})}
\newcommand{\refeqs}[2]{Eqs.~(\ref{eq:#1})-(\ref{eq:#2})}
\newcommand{\reffig}[1]{Fig.~\ref{fig:#1}}
\newcommand{\refsec}[1]{Section \ref{sec:#1}}
\newcommand{\reftab}[1]{Table \ref{tab:#1}}
\newcommand{\order}[1]{\mathcal{O}\({#1}\)}
\newcommand{\fv}[1]{\left(\begin{array}{c}#1\end{array}\right)}%

\def\tcb#1{\textcolor{blue}{#1}}
\def\tcr#1{\textcolor{red}{#1}}
\def\tcg#1{\textcolor{green}{#1}}
\def\tcc#1{\textcolor{cyan}{#1}}
\def\tcv#1{\textcolor{violet}{#1}}
\def\tcm#1{\textcolor{magenta}{#1}}
\def\tcpn#1{\textcolor{pink}{#1}}
\def\tcpr#1{\textcolor{purple}{#1}}
\definecolor{schrift}{RGB}{120,0,0}

\def \azeL{{H_0^L}}
\def \azeR{{H_0^R}}
\def \apaL{{H_\parallel^L}}
\def \apaR{{H_\parallel^R}}
\def \apeL{{H_\perp^L}}
\def \apeR{{H_\perp^R}}

%% Physics
\newcommand{\alphas}{\alpha_\mathrm{s}}
\newcommand{\alphae}{\alpha_\mathrm{e}}
\newcommand{\gfermi}{G_\mathrm{F}}
\newcommand{\GeV}{\,\mathrm{GeV}}
\newcommand{\MeV}{\,\mathrm{MeV}}
\newcommand{\amp}[1]{\mathcal{A}\left({#1}\right)}
\newcommand{\wilson}[2][{}]{\mathcal{C}_{#2}^{\mathrm{#1}}}
\newcommand{\bra}[1]{\left\langle{#1}\right\vert}
\newcommand{\ket}[1]{\left\vert{#1}\right\rangle}

%%%%
%    Greek Letters
%

\let\a=\alpha      \let\b=\beta       \let\c=\chi        \let\d=\delta
\let\e=\varepsilon \let\f=\varphi     \let\g=\gamma      \let\h=\eta
\let\k=\kappa      \let\l=\lambda     \let\m=\mu
\let\o=\omega      \let\r=\varrho     \let\s=\sigma
\let\t=\tau        \let\th=\vartheta  \let\y=\upsilon    \let\x=\xi
\let\z=\zeta       \let\io=\iota      \let\vp=\varpi     \let\ro=\rho
\let\ph=\phi       \let\ep=\epsilon   \let\te=\theta
\let\n=\nu
\let\D=\Delta   \let\F=\Phi    \let\G=\Gamma  \let\L=\Lambda
\let\O=\Omega   \let\P=\Pi     \let\Ps=\Psi   \let\Si=\Sigma
\let\Th=\Theta  \let\X=\Xi     \let\Y=\Upsilon
%
%%%

%%%
%    Calligraphic letters

\def\cA{{\cal A}}                \def\cB{{\cal B}}
\def\cC{{\cal C}}                \def\cD{{\cal D}}
\def\cE{{\cal E}}                \def\cF{{\cal F}}
\def\cG{{\cal G}}                \def\cH{{\cal H}}
\def\cI{{\cal I}}                \def\cJ{{\cal J}}
\def\cK{{\cal K}}                \def\cL{{\cal L}}
\def\cM{{\cal M}}                \def\cN{{\cal N}}
\def\cO{{\cal O}}                \def\cP{{\cal P}}
\def\cQ{{\cal Q}}                \def\cR{{\cal R}}
\def\cS{{\cal S}}                \def\cT{{\cal T}}
\def\cU{{\cal U}}                \def\cV{{\cal V}}
\def\cW{{\cal W}}                \def\cX{{\cal X}}
\def\cY{{\cal Y}}                \def\cZ{{\cal Z}}

%%%%

\def\be{\begin{equation}}
\def\ee{\end{equation}}
\def\bea{\begin{eqnarray}}
\def\eea{\end{eqnarray}}
\def\bm{\begin{matrix}}
\def\em{\end{matrix}}
\def\bpm{\begin{pmatrix}}
    \def\epm{\end{pmatrix}}

{\newcommand{\lsim}{\mbox{\raisebox{-.6ex}{~$\stackrel{<}{\sim}$~}}}
{\newcommand{\gsim}{\mbox{\raisebox{-.6ex}{~$\stackrel{>}{\sim}$~}}}
\def\mpl{M_{\rm {Pl}}}
\def\gev{{\rm \,Ge\kern-0.125em V}}
\def\tev{{\rm \,Te\kern-0.125em V}}
\def\mev{{\rm \,Me\kern-0.125em V}}
\def\ev{\,{\rm eV}}

\title{\boldmath  Towards a unified explanation of $R_{D^{(\ast)}}$, $R_{K}$ and $(g-2)_{\mu}$ anomalies in a left-right model with leptoquarks}
\author{Diganta Das}
\email{diganta@prl.res.in} 
\affiliation{Physical Research Laboratory, Navrangpura, Ahmedabad 380 009, India}
\author{Chandan Hati}
\email{chandan@prl.res.in} 
\affiliation{Physical Research Laboratory, Navrangpura, Ahmedabad 380 009, India}
\affiliation{Indian Institute of Technology Gandhinagar, Chandkheda, Ahmedabad 382 424, India}
\author{Girish Kumar}
\email{girishk@prl.res.in}
\affiliation{Physical Research Laboratory, Navrangpura, Ahmedabad 380 009, India}
\affiliation{Indian Institute of Technology Gandhinagar, Chandkheda, Ahmedabad 382 424, India}
\author{Namit Mahajan}
\email{nmahajan@prl.res.in} 
\affiliation{Physical Research Laboratory, Navrangpura, Ahmedabad 380 009, India}

\begin{abstract}
We present a unified explanation for the $B$-decay anomalies in $R_{D^{(\ast)}}$ and $R_{K}$ together with the anomalous muon magnetic moment, consistent with the constraints from the current measurements of leptonic decay rates and $D^0-\bar{D}^0$, $B_{s}^0-\bar{B}_{s}^0$ mixings, within the framework of a minimal left-right symmetric gauge theory motivated by one of the low-energy subgroups of $E_{6}$ naturally accommodating leptoquarks.
	
\end{abstract}
%\pacs{98.80.Cq,12.60.-i}
%Particle-theory models (Early Universe), 98.80.Cq
%models beyond the standard models, 12.60.-i
\maketitle
%%%%%%%%%%%%%%%%%%%%%%%%%%%%%%%%%%%%%%%%%%%%%%%%%%%%%%%%%%%%%%%%%%%%%%%%%%%%%%%%%%%%%%%%%%%%%%%%%%%%%%%%%%%%%%%%%%%%%%%%%%%%%%%%%%%%%%%%%%%%%%%%%%%%%%%%
{\uppercase\expandafter{\romannumeral 1.\relax} {\bf{Introduction ---}}
Precision measurements associated with rare decays provide powerful probes for new physics (NP) beyond the Standard Model (SM) in the intensity frontier of modern particle physics. To this end, the study of rare $B$ decays induced by  flavor changing neutral current (FCNC) have shown some interesting anomalies hinting towards lepton nonuniversal NP. In 2012, the BABAR Collaboration reported \cite{Lees:2012xj} the measurements of the ratio of branching fractions 
\be
R_{D^{(\ast)}}=\frac{{\rm{Br}}(\bar{B}\rightarrow D^{(\ast)}\tau\bar{\nu})}{{\rm{Br}}(\bar{B}\rightarrow D^{(\ast)} l \bar{\nu})},
\ee
$R_D^{{\rm BABAR}}=0.440\pm 0.058  \pm 0.042 $ and $R_{D^{\ast}}^{{\rm BABAR}}=0.332\pm 0.024 \pm 0.018 $ showing $2.0 \sigma$ and $2.7\sigma$ enhancements over the SM predictions $R_D^{{\rm SM}}=0.300\pm 0.010$ and $R_{D^{\ast}}^{{\rm SM}}=0.252\pm 0.005$, respectively. Partially corroborating this result in 2015, the Belle Collaboration reported $R_D^{{\rm Belle}}=0.375\pm 0.064  \pm 0.026 $ and $R_{D^{\ast}}^{{\rm Belle}}=0.293\pm 0.038  \pm 0.015 $ \cite{Huschle:2015rga}. Very recently, the LHCb and Belle Collaborations have reported $R_{D^{\ast}}^{{\rm LHCb}}=0.336\pm 0.027 ({\rm stat.}) \pm 0.030 ({\rm syst.})$ and $R_{D^{\ast}}^{{\rm Belle16}}=0.302\pm 0.030 ({\rm stat.}) \pm 0.011 ({\rm syst.})$ amounting to $\sim 2.1\sigma$ and $\sim 1.6\sigma$ enhancements over the SM predictions, respectively \cite{Aaij:2015yra, Abdesselam:2016cgx}. These results are consistent with each other and when combined together show significant enhancements over the SM expectations, hinting towards a large new physics 
contribution. Interestingly, the LHCb Collaboration \cite{Aaij:2014ora} has recently reported another striking deviation from the SM prediction of the ratio of branching fractions of charged $\bar{B}\rightarrow \bar{K}ll$ decays
\be
R_{K}=\frac{{\rm{Br}}(\bar{B}\rightarrow \bar{K}\mu^{+}\mu^{-})}{{\rm{Br}}(\bar{B}\rightarrow \bar{K} e^{+} e^{-})}.
\ee
The measured value of $R_{K}^{\rm{LHCb}}=0.745\pm_{0.074}^{0.090}\pm 0.036$, in the dilepton invariant mass squared bin $1 \gev^{2}\leq q^{2}\leq 6  \gev^{2}$ corresponds to a $2.6 \sigma$ deviation from the SM prediction $R_{K}^{\rm{SM}}=1.0003\pm0.0001$ \cite{Bobeth:2007dw}.

On the other hand, currently the most precise measurement of the anomalous muon magnetic moment by E821 experiment at Brookhaven National Laboratory (BNL) has been reported to show a significant deviation from the SM prediction $\Delta a_{\mu}=a_{\mu}^{\rm{exp}}-a_{\mu}^{\rm{SM}}=(2.8 \pm 0.9)\times 10^{-9}$ amounting to a $\sim 3\sigma$ level deviation \cite{Bennett:2004pv}. This discrepancy also points to the possible existence of NP beyond the SM. 

Several attempts have been made in the literature to explain the above anomalies in $B$ decays by invoking NP models \cite{Bhattacharya:2014wla,Glashow:2014iga,Alonso:2015sja,Greljo:2015mma,Calibbi:2015kma,Bauer:2015knc,Fajfer:2015ycq,Boucenna:2016wpr,Belanger:2015nma,Celis:2012dk} and separately using a model independent approach \cite{Sakaki:2013bfa,Alonso:2014csa,Hiller:2014yaa,Hurth:2014vma,Couture:1995he,Dorsner:2016wpm,Datta:2012qk} \footnote{Some works also try to address 750GeV diphoton excess together with these anomalies \cite{Bauer:2015boy}.}. Among them, one of the extensively studied class of models relies on scalar or vector leptoquarks. However, in these models the leptoquark couplings are often taken at an effective level without any concrete framework. The purpose of this paper is to explain all three anomalies consistently within the framework of a left-right symmetric gauge theory naturally accommodating leptoquarks. This framework, motivated by one of the low energy subgroups of $E_6$, 
can naturally enhance both $\bar{B}\rightarrow D \tau \bar{\nu}$ and $\bar{B}\rightarrow D^{(\ast)}\tau \bar{\nu}$ via the exchange of scalar leptoquarks to explain the anomalies, while the $R_K$ data can be explained simultaneously through one loop diagrams induced by leptoquarks. The anomalous muon magnetic moment can also be explained in this model without utilizing a nonzero right-handed coupling of leptoquarks. We also discuss various constraints from the current measurements of (semi) leptonic decays and $B_{s}^0-\bar{B}_{s}^0$, $D^0-\bar{D}^0$ mixings and comment on nonzero branching fraction of the lepton flavor violating decay $h\to\tau\mu$ which requires a fine-tuning to satisfy constraints from $\tau\to\mu\gamma$ decay. 
%%%%%%%%%%%%%%%%%%%%%%%%%%%%%%%%%%%%%%%%%%%%%%%%%%%%%%%%%%%%%%%%%%

{\uppercase\expandafter{\romannumeral 2.\relax} \bf{Low energy subgroups of $E_6$ and Neutral Left-Right Symmetric Model ---}}
 Superstring inspired $E_6$ grand unified theory (GUT) being the next natural anomaly free GUT theory after $SO(10)$ has obtained considerable attention in literature thanks to many alternative intermediate mass breaking scales and a number of exotic new fields including leptoquarks or diquarks (to avoid rapid proton decay both cannot be present simultaneously) promising a rich low energy phenomenology. One of the maximal subgroups of $E_6$ is given by $SU(3)_c\times SU(3)_L\times SU(3)_R$. Under this subgroup, the fundamental 27 representation of $E_6$ has the decomposition given by
\begin{equation}\label{eq:27rep}
 27 = (3,3,1) + (3^*,1,3^*) + (1,3^*,3)\, .
\end{equation}
The assignment of the multiplets are as follows $(u,d,h): (3,3,1)$, $(h^c,d^c,u^c):(3^*,1,3^*)$,  and the leptons are assigned to $(1,3^*,3)$. 
Other than the usual SM fields, several new fields are present including an exotic $-1/3$ charge leptoquark $h$, the right-handed neutrino $N^c$ and the two lepton isodoublets $(\nu_E, E)$ and $(E^c,N^c_E)$. The breaking of $SU(3)_{L}$ to $SU(2)_{L}\times U(1)_{L}$ is fixed by the SM isodoublet structure, for example, $(u,d,h)_{L}: (3,3,1)$ must break to the usual SM isodoublet $(u,d)_{L}$ and an isosinglet $h_{L}$. However, there are three choices to break $SU(3)_{R}$ to $SU(2)_{R}\times U(1)_{R}$ depending on the three possible choices of the $SU(2)_R$ doublet corresponding to $T, U, V$ isospins of $SU(3)_{R}$. The three choices of the residual $SU(2)_{R}$ give three possible left-right symmetric frameworks. In this paper, we are interested in the choice where $(h^c,d^c)_{L}$ is the residual $SU(2)_R$ isodoublet \cite{London:1986dk}. Interestingly, this choice results in a unique situation where the residual $SU(2)_{R}$ does not contribute to the electric charge \cite{London:1986dk}, and hence, we call this 
model ``neutral" left-right symmetric model (NLRSM). We will denote the residual $SU(2)_{R}$ as $SU(2)_N$. The corresponding charge equation is given by $Q = Y_{3L} + \frac{1}{2}Y_L + \frac{1}{2}Y_N$. The fields have the following transformations under the NLRSM gauge group $G = SU(3)_c\times SU(2)_L\times SU(2)_N\times U(1)_Y$ \cite{Dhuria:2015hta}
\begin{eqnarray}\label{eq:mltplt}
(u,d)_L&:& (3,2,1,\frac{1}{6}), \quad (h^c,d^c)_L: (\bar{3},1,2,\frac{1}{3}) ,\nn\\
  (E^c,N_E^c)_L&:& (1,2,1,\frac{1}{2}),\quad (N^c,n)_L: (1,1,2,0),\nn\\
   h_L&:& (3,1,1,-\frac{1}{3}), \quad \quad \quad u_L^c: (\bar{3},1,1,-\frac{2}{3}),\nn\\
e_L^c&:& (1,1,1,1), \quad \begin{pmatrix} \nu_e & \nu_E \\ e & E \end{pmatrix} : (1,2,2,-\frac{1}{2}). 
\end{eqnarray}
The gauge bosons corresponding to $SU(2)_N$ are electrically neutral and are denoted by $Z_N, W_N^\pm$, where the $\pm$ sign refers to $SU(2)_N$ charge. The interactions of the new exotic fields with the SM sector are governed by the superpotential 
 \begin{eqnarray}
 \label{eq:Wcase2}
 && {\hskip -0.2in} W= \lambda^1 \left( \nu_e N^{c}_L N^{c}_E + e E^c N^{c}_L +\nu_E N^{c}_E n + E E^c n \right) \nonumber\\
 &&  {\hskip -0.2in}+ \lambda^2 \left( d^c N^{c}_L h+ h h^c n \right) +\lambda^3 u^c e^c h+ \lambda^4 \left( u u^c N^{c}_E + u^c d E^c\right) \nonumber\\
 &&  {\hskip -0.2in}+ \lambda^5 \left(\nu_e e^c E+ e e^c \nu_E \right) +\lambda^6 \left(u d^c E + d d^c \nu_E + u h^c e + d h^c \nu_e \right). \nonumber\\
  \end{eqnarray}
From the superpotential, it follows that the leptoquark $h$ has the assignment $B=1/3$ and $L=1$, while the exotic fields $\nu_E, E$, and $n$ have $B=L=0$ and $N^c$ has $B=0, L=-1$. In the gauge sector, $W_N$ carries a nonzero lepton number $B=0, L=-1$. 

In addition to the above superpotential couplings, the gauge couplings of $W_{N}$ and $Z_{N}$ to the fermions can also induce FCNC processes such as $B^0-\bar{B}^0, K^0-\bar{K}^0$ and in the leptonic sector, the lepton flavor violating (LFV) processes such as the decays $h\to\tau\mu$, $\mu\to e\gamma$, as well as can contribute to the anomalous muon magnetic moment in the presence of mixing between new exotic fields  \cite{Hewett:1988xc}. To keep things minimal, in the following analysis, we assume that the dominant FCNC and LFV contributions come from scalar leptoquark induced processes \footnote{A more complex situation where such mixings are present and $W_{N}$ induced processes can contribute significantly will be discussed in a separate communication. Note that, among other $E_{6}$ low energy subgroups Alternative Left-Right Symmetric Model \cite{Ma:1986we} and variants of $U(1)_{N}$ model \cite{Dhuria:2015swa} have leptoquark couplings somewhat similar to this model. Also, $R$-parity violating supersymmetry model can be one of the interesting candidates to explain these anomalies and at an effective level can be somewhat similar to this model \cite{Deshpande:2012rr,Biswas:2014gga}}.

At the LHC, the pair production of scalar leptoquarks is studied through the decay of leptoquarks into quark-lepton pairs $eq$ or $\nu q$ [see the interactions in Eq. (\ref{eq:Wcase2})], which give bounds on the scalar leptoquark mass almost independent of the coupling coefficient. The current limits on leptoquark masses from CMS searches for pair production of scalar leptoquarks are  830, 840, and 525 GeV for first, second, and third generations, respectively. From ATLAS searches these limits are 660, 422, and 534 GeV, respectively \cite{Agashe:2014kda}. On the other hand, from single production search, the current lower limit on the first generation scalar leptoquark is 304 GeV \cite{Agashe:2014kda}. We have chosen the benchmark LQ mass as an average of the most stringent limits from CMS and ATLAS; nevertheless, for a higher leptoquark mass, all the conclusions remain equally valid. In passing, we would also like to mention that the existence of heavy neutral gauge bosons and new exotic fermions can induce shifts to the electroweak precision observables. However, a sufficiently small mixing between heavy $Z^{\prime}$ and the SM $Z$ ensures that this shift is negligible \cite{Hewett:1988xc}. The exotic fermions can contribute to a shift to a $\rho$ parameter via one-loop vacuum polarization diagrams; however, given that the exotic states are vectorlike particles, one must perform an analysis for general vector and axial-vector couplings. This is beyond the scope of the present work and can be found in Ref. \cite{Rizzo:1986ws}.
%%%%%%%%%%%%%%%%%%%%%%%%%%%%%%%%%%%%%%%%%%%%%%%%%%%%%%%%%%%%%%%%%%%%%%%%%%%%%%%%%%%%%%%%%%%%%%%%%%%%%%%%%%%%%%%%%%%%%%%%%%%%%%%%%%%%%%%%%%%%%%%%%%%%%%%%%%%%%%%%%%%%%%%%%%%%

{\uppercase\expandafter{\romannumeral 3.\relax} \bf{Explaining $R_{D^{(\ast)}}$ anomalies and constraints from $D^0-\bar{D}^0$ mixing and $B$, $D$ decays---}}
In NLRSM, the scalar leptoquark ($\tilde{h}^{\ast}$) and slepton ($\tilde{E}$) can mediate the semileptonic decays $\bar{B}\rightarrow D^{(\ast)}\tau \bar{\nu}$ at tree level. The effective Lagrangian is given by
\bea
\label{3.1}
\cL_{\rm{eff}}=\sum_{i,k=1}^{3} V_{2i}\left[ \frac{ \l^{5}_{33k} \l^{6\ast}_{i3k}}{m_{\tilde{E}^k}^{2}} \bar{c}_{L}b_{R} \; \bar{\tau}_{R}\nu_{L}\right. \nonumber\\
\left.+ \frac{ \l^{6}_{33k} \l^{6\ast}_{i3k}}{m_{\tilde{h}^{k \ast}}^{2}} \bar{c}_{L}(\tau^{c})_{R} \; (\bar{\nu}^{c})_{R} b_{L} \right],
\eea
where the superscripts are superpotential coupling indices and the generation indices are written as subscripts. Here, $m_{\tilde{h}} (m_{\tilde{E}})$ is the mass of scalar leptoquark $\tilde{h}^{k\ast}$ (slepton $\tilde{E}^{k}$) and $V_{ij}$ is the $ij$th component of the CKM matrix. In the convention of Ref. \cite{Hati:2015awg}, the Wilson coefficients are given by
\bea
\label{3.3}
C^{\tau}_{S_L} &=& -\frac{1}{2\sqrt{2} G_{F} V_{cb}} \sum_{i,k=1}^{3} V_{2i} \frac{ \l^{5}_{33k} \l^{6\ast}_{i3k}}{m_{\tilde{E}^k}^{2}} \; ,\nonumber\\
C^{\tau}_{V_L} &=& -\frac{1}{2\sqrt{2} G_{F} V_{cb}} \sum_{i,k=1}^{3} V_{2i} \frac{ \l^{6}_{33k} \l^{6\ast}_{i3k}}{2\,m_{\tilde{h}^{k \ast}}^{2}},
\eea
where the neutrinos are assumed to be of a tau flavor. To simplify further analysis, we assume that except the SM contribution only the scalar leptoquark NP operator contributes dominantly. This is justified because, the case where $C^{\tau}_{S_L}$ is the dominant contribution, similar to 2HDM of type II or type III with minimal flavor violation, can not explain both $R_D$ and $R_{D^{\ast}}$ data simultaneously \cite{Sakaki:2013bfa, Crivellin:2012ye}.
    
The leptonic decay modes $B \rightarrow \tau \nu $, $D_s^+ \rightarrow \tau \nu $, $D^+ \rightarrow \tau \nu $, and $D^0-\bar{D}^0$ mixing induced by scalar leptoquark $\tilde{h}^{k\ast}$ exchange can be utilized to derive constraints on the product of  couplings $\l^{6}_{33k} \l^{6 \ast}_{13k}$ using measured branching fractions for the decays and $D^0-\bar{D}^0$ mixing parameters. In NLRSM, the exchange of the scalar leptoquark $\tilde{h}^{k\ast}$ leads to an additional tree level diagram for the decay $B \rightarrow \tau \nu $ in addition to the usual SM contribution. Assuming couplings to be real, the modified rate of the decay process $B \rightarrow \tau \nu $ gives constraint on the product of couplings $\l^{6}_{33k} \l^{6\ast}_{13k}$ given by
 \bea
\label{4.1.3}
- 0.04 \left( \frac{m_{\tilde{h}^{k \ast}}}{1000 \rm{GeV}}\right)^2 \leq \l^{6}_{33k} \l^{6}_{13k}\leq 0.03 \left( \frac{m_{\tilde{h}^{k \ast}}}{1000 \rm{GeV}}\right)^2 .
\eea
The measured branching ratios of the decays $D_s^+ \rightarrow \tau \nu $ and $D^+ \rightarrow \tau \nu $ can be used to obtain constraints on $(\l_{23k}^{6})^2$ and $\l_{23k}^{6} \l_{13k}^{6}$, respectively.  The decay $D_s^+ \rightarrow \tau \nu $ gives the constraint
 \bea
 \label{4.2.3}
(\l_{23k}^{6})^2 \leq 1.9 \left( \frac{m_{\tilde{h}^{k \ast}}}{1000 \rm{GeV}}\right)^2 ,
 \eea
and the decay process $D^+ \rightarrow \tau \nu $ gives a weaker constraint on $\l_{23k}^{6} \l_{13k}^{6}$ compared to $D^{0}-\bar{D}^{0}$. The relevant box diagrams are similar to the diagrams generated from internal line exchange of lepton-squark pair or slepton-quark pair in the case of $R$-parity violating models \cite{Agashe:1995qm, Golowich:2007ka}. The relevant constraint is given by
  \bea
\label{4.2.2}
-0.012 \left(\frac{m_{\tilde{h}^{k \ast}}}{1000 \rm{GeV}}\right) \leq \l_{23k}^{6} \l_{13k}^{6} \leq 0.012 \left(\frac{m_{\tilde{h}^{k \ast}}}{1000 \rm{GeV}}\right).
\eea
\begin{figure}[ht!]
   \hspace{0.02cm}
    \hbox{\hspace{0.03cm}
    \hbox{\includegraphics[scale=0.35]{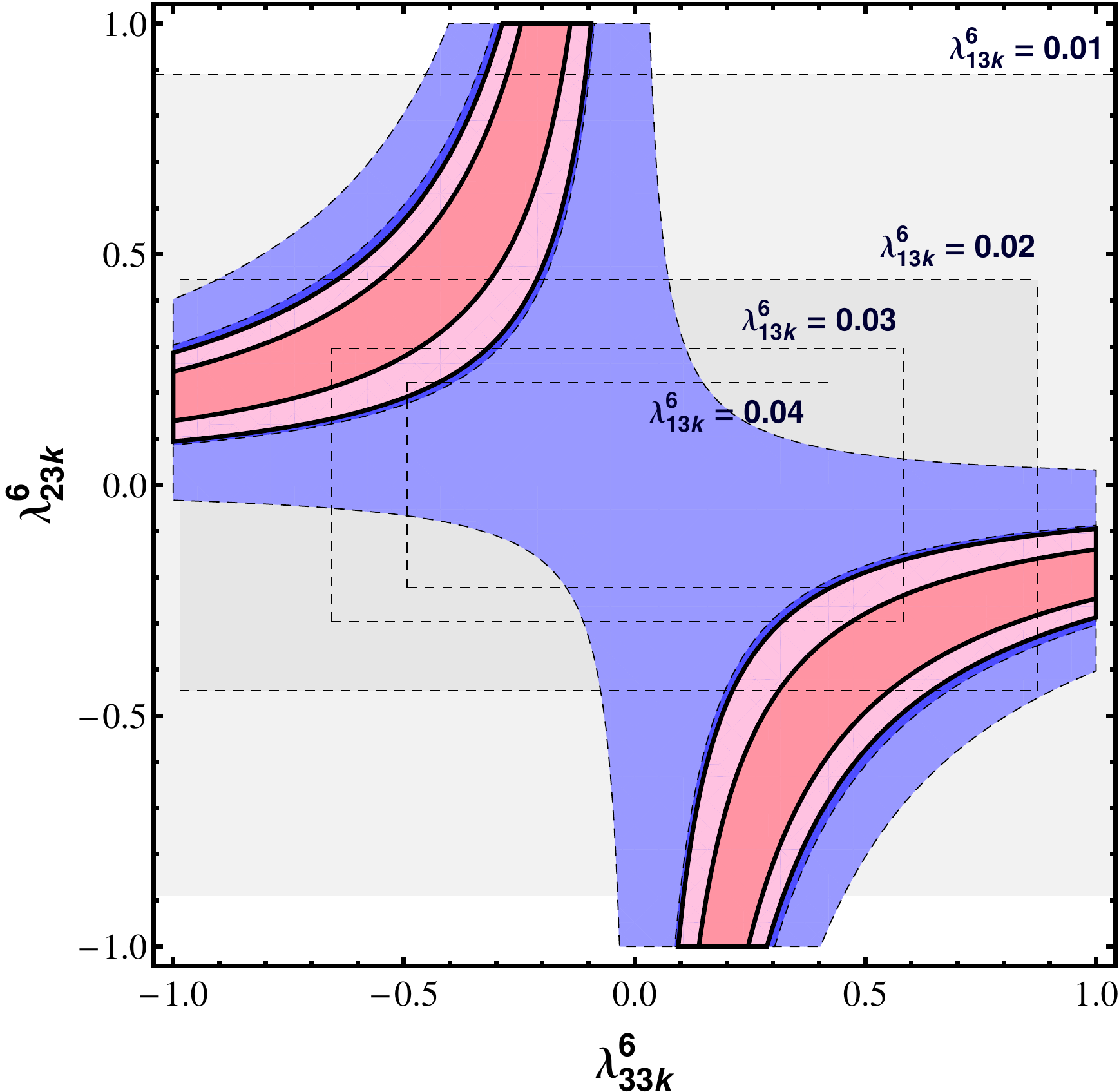}}
    }
      \caption{ $R_{D^{(\ast)}}$  compatible $\lambda_{33k}^{6}-\lambda_{23k}^{6}$ parameter space constrained from $B \rightarrow \tau \nu $, $D_s^+ \rightarrow \tau \nu$, and $D^0-\bar{D}^0$ mixing. For details, see text.}
     \label{fig:RD}
    \end{figure}

In Fig. \ref{fig:RD}, we plot the range of the couplings $\l_{33k}^{6}$ and $\l_{23k}^{6}$ (for $m_{\tilde{h}^k\ast}=750 \gev$) that can explain both $R_D$ and $R_{D^{\ast}}$ data over the parameter space allowed by the leptonic decays and $D^{0}$-$\bar{D}^{0}$ mixing. The shaded (light gray) rectangles with dashed boundaries correspond to regions of $\l_{33k}^{6}$-$\l_{23k}^{6}$ parameter space allowed by the constraints from the $B \rightarrow \tau \nu $, $D_s^+ \rightarrow \tau \nu $ decays and $D^0-\bar{D}^0$ mixing for different values of $\l_{13k}^{6}$. The solid (deep) blue bands correspond to the $(1\sigma) 2\sigma$ allowed band explaining the $R_D$ data and the (deep) pink bands correspond to the allowed band explaining both $R_D$ and $R_{D^{\ast}}$ data simultaneously. Note that the list of constraints mentioned above is not complete and more independent constraints can be derived from other leptonic and semileptonic decays. For example, the decay processes $\tau^{+} \rightarrow \pi^{+} \nu$ and 
$t \rightarrow b \tau \nu$ can give independent constraint on $\l_{13k}^{6}$ and $\l_{33k}^{6}$,  respectively, which we find to be consistent with the ones discussed above. Finally, the effective NP operators under consideration also predict an enhanced decay rate for $b\rightarrow s\nu \bar{\nu}$ \cite{Grossman:1995gt, Buras:2014fpa}, which can be an interesting channel for the future experiments and can be intriguing in the context of radiative neutrino masses.

%%%%%%%%%%%%%%%%%%%%%%%%%%%%%%%%%%%%%%%%%%%%%%%%%%%%%%%%%%%%%%%%%%%%%%%%%%%%%%%%%%%%%%%%%%%%%%%%%%%%%%%%%%%%%%%%%%%%%%%%%%%%%%%%%%%%%%%%%%%%%%%%%%%%%%%%%%%%%%%%%%%%%%%%%%%%%%%%%%%%%%%%

{\uppercase\expandafter{\romannumeral 4.\relax} \bf{Explaining $R_K$ anomaly and constraints from $B_{s}^0-\bar{B}_{s}^0$ mixing---}}
The lepton nonuniversality in the ratio $R_K$ has been analyzed in a model-independent fashion in Refs. \cite{Alonso:2014csa, Hiller:2014yaa} suggesting that a good fit to the data is obtained for the constraints
\begin{eqnarray} \label{RK:con}
-1.5 \lesssim &C^{\mu}_{LL} &\lesssim -0.7\, , \nonumber\\
-1.9\lesssim &C^{\mu}_{LL}-C^{\mu}_{LR}&\lesssim 0.
\end{eqnarray}
The study  \cite{Hiller:2014yaa} has also discussed leptoquark induced tree level contributions which require either very large leptoquark masses or small couplings in order to explain the data. In Ref. \cite{Bauer:2015knc} it was explicitly pointed out that one loop box diagrams can also explain the departure from the SM prediction for $\mathcal{O}(1)$ left-handed couplings and suppressed right-handed couplings. In NLRSM also, the $b\to s\ell\ell$ flavor changing transition can occur at one loop level via the scalar and fermionic leptoquarks induced box diagrams shown in Fig.~\ref{fig:btos}. We find that the $\gamma$- and $Z$-penguin diagrams (including their supersymmetric counterparts) give a vanishing contribution
\begin{figure}[h!]
\begin{center}
 \includegraphics[scale=0.5]{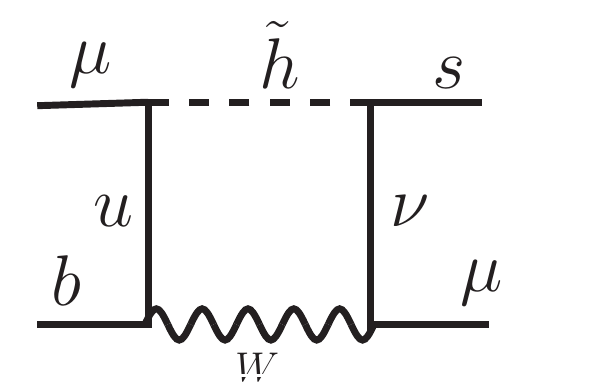}
 \includegraphics[scale=0.5]{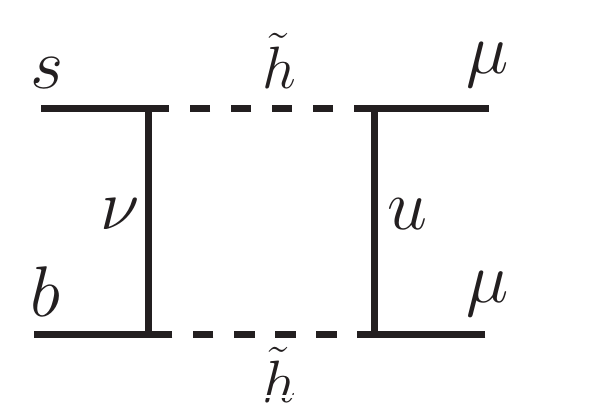}
 \includegraphics[scale=0.35]{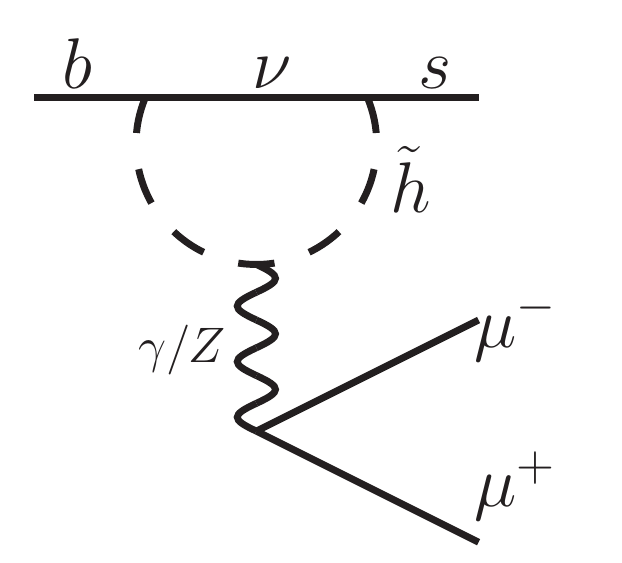}
 \caption{Representative diagrams for $b\to s\ell\ell$ transition. }
    \label{fig:btos}
\end{center}
 \end{figure}
  \footnote{ Note that that a careful dimensional reduction ensures a vanishing contribution from Z-penguin diagrams Ref. \cite{Lunghi:1999uk,Abada:2012cq,Krauss:2013gya,Abada:2014kba}. Note that there are also diagrams involving charginos and neutralinos at one loop level independent of the leptoquarks giving a subdominant contribution.}, which is in agreement with Ref. \cite{Bauer:2015knc}.
The  contribution to  $C^{\mu}_{ LL }$ from scalar as well as fermionic leptoquark induced box diagrams in the limit $m^{2}_{\tilde{h},h}\gg m^{2}_{W,t}$  is given by
\begin{eqnarray}{\label{RKWC}}
C^{\mu}_{ LL }&=&\frac{\lambda_{32k}^{6}\lambda^{6\ast}_{32k} }{8\pi \a_e} \left(\frac{m_t}{m_{\tilde{h}_j}}\right)^2 - \frac{\lambda_{3jk}^{6}\lambda^{6\ast}_{2jl}\lambda_{i2k}^{6}\lambda^{6\ast}_{i2l}}{32\sqrt{2}\, G_F\, V_{tb}V_{ts}^{\ast} \pi \alpha_e m^2_{\tilde{h}} }  \nonumber\\
&-& \frac{\lambda_{3jk}^{6}\lambda^{6\ast}_{2jl}\lambda_{i2k}^{6}\lambda^{6\ast}_{i2l}}{32\sqrt{2}\, G_F\, V_{tb}V_{ts}^{\ast} \pi \alpha_e m^{2}_h }\,\, g\!\left(\frac{m_{\tilde{u}_i}^2}{m_h^2},1,\frac{m^2_{\tilde{\nu}_j}}{m_h^2} \right),
\end{eqnarray} 
 where repeated indices are summed over and the loop function $g(x,y,z)$ is defined by 
  \begin{eqnarray}{\label{loopfnt}}
  &&g(x,y,z) = \frac{x^2 \log x}{(x-1)(x-y)(x-z)} + (\rm{cycl.\,\,\,  perm.}).\nonumber
  \end{eqnarray} 
Note that $C^{\mu}_{ LL }$ depends on the product of couplings $\lambda_{3jk}^{6}\lambda^{6\ast}_{2jk}$ with the $j=3$ set of couplings appearing in the Wilson coefficient $C^{\tau}_{V_L}$ in Eq. (\ref{3.3}). The contribution from the box diagrams also involves one additional set of couplings $\lambda^{6\ast}_{i2k}\lambda_{i2l}^{6}$ which can be constrained from the measurement of $Z\to\mu\bar{\mu}$ decay rate. In Ref. \cite{Bauer:2015knc}, it was found that for a $\tev$  scale leptoquark, the size of such couplings can be as large as $\sim \mathcal{O}(1)$. Processes such as $t \rightarrow b \mu \bar{\nu}_\mu$, $D_s \rightarrow \mu \bar{\nu}_\mu$ etc give similar constraints on individual couplings $\lambda^{6\ast}_{i2k}\lambda_{i2l}^{6}$. The product of couplings $\lambda_{3jk}^{6}\lambda^{6\ast}_{2jk}$ contributes to $B_{s}-\bar{B}_{s}$ mixing amplitude. Following the suggestion of the UT$fit$ Collaboration \cite{Bona:2007vi}, we 
define the ratio $C_{B_s} e^{2 i \phi_{B_s} } = \langle B_s |H^\text{full}_\text{eff} | \bar{B}_s \rangle / \langle B_s |H^\text{SM}_\text{eff} | \bar{B}_s \rangle $ to obtain 
\begin{equation}
 C_{B_s}e^{2i\phi_{B_s}} = 1 + \frac{m_W^2}{g^4 S_0(x_t)} \Big( \frac{1}{m^2_{\tilde{h}}} + \frac{1}{m^2_h} \Big) \frac{\lambda_{3jk}^{6} \lambda_{3lm}^{6} \lambda_{2jm}^{6\ast} \lambda_{2lk}^{6\ast}}{( {V_{tv} V^{\ast}_{ts} })^2},
\end{equation}
which gives an allowed range consistent with the value of $\lambda^{6}_{3jk}\lambda^{6\ast}_{2jk}$ required to explain 
the $R_{K}$ data using the latest UT$fit$ values of the $B_{s}-\bar{B}_{s}$ mixing parameters. As a benchmark point 
taking $\lambda_{3jk}^{6}\lambda^{6\ast}_{2jk} \simeq 0.07$ for $m_{\tilde{h}}\sim 750$ GeV, $m_{h}\sim 600$ GeV and 
taking $\lambda^{6\ast}_{i2k} \sim \mathcal{O}(1)$, we obtain the standard benchmark solution $C^{\mu}_{ LL }=-1$ and $C^{\mu}_{ LR}=0$ which 
satisfies the conditions given in Eq. (\ref{RK:con}). 
The leptoquarks give additional contributions to $b\to s\gamma$ transition through penguin diagrams involving $\tilde{\nu}-h$ and $\nu-\tilde{h}$ in loops. 
Explicitly calculating the diagrams we find that the contributions to $C_{7\gamma}$ is negligibly small, which is in agreement with Ref. \cite{Bauer:2015knc}.

%%%%%%%%%%%%%%%%%%%%%%%%%%%%%%%%%%%%%%%%%%%%%%%%%%%%%%%%%%%%%%%%%%%%%%%%%%%%%%%%%%%%%%%%%%%%%%%%%%%%%%%%%%%%%%%%%%%%%%%%%%%%%%%%%%%%%%%%%%%%%%%%%%%%%%%%%%%%%%%%%%%%%%%%%%%%%%%%%%%%%%%%

{\uppercase\expandafter{\romannumeral 5.\relax} \bf{Explaining anomalous muon magnetic moment ---}}
In the SM, the muon anomalous magnetic moment is chirally suppressed due to a small muon mass, $a_\mu\sim m_\mu^2/m_{W}^2$. In NLRSM, leptoquarks can induce an additional contribution to the anomalous magnetic moment of the muon through one-loop vertex diagrams. However, the sole contribution from leptoquark induced diagrams cannot explain the experimental deviation from the SM. One way out is to follow the approach taken in Ref. \cite{Bauer:2015knc}, where a nonzero right-handed coupling of a leptoquark is utilized. Interestingly, in NLRSM, it is possible to explain the experimental data through a dominant contribution from $\lambda^{5}$ terms in Eq. (\ref{eq:Wcase2}). The new contribution from $\l^6_{ijk}$ terms in Eq. (\ref{eq:Wcase2}) is given by 
\bea
a_\mu (\l^6) &=& \frac{m_\mu^2}{32\pi^2}\left[\frac{1}{m^2_{\tilde{h}_{jR}^{*}}-m_t^2}\lvert \l^6_{32j}\rvert^2 \left( 1+\frac{2 x_t}{1-x_t}\right)\right.\nonumber\\
&& \left.\left(\frac{1}{2}+\frac{3}{1-x_t}+\frac{2+x_t}{(1-x_t)^2} \ln \,x_t\right) \right],
\eea
where $x_t=m_{t}^{2}/m_{\tilde{h}}^{2}$. The $\l^5_{ijk}$ terms in Eq. (\ref{eq:Wcase2}) give the following contribution induced by sleptons
\bea
\delta a_\mu(\l^5) &=& \frac{m_\mu^2}{16\pi^2}  \left[ \lvert \l_{i2k}\rvert^2 F(e_k,\tilde{\nu}_{Ei}) -\lvert \l_{i2k}\rvert^2 F(\tilde{e}_k,\nu_{Ei}) \right.\nonumber\\
&& + \lvert \l_{ij2}\rvert^2 F(e_j,\tilde{\nu}_{Ei} )- \lvert \l_{ij2}\rvert^2 F(\tilde{e}_{j},\nu_{Ei})  \nonumber\\
&& \left. + \lvert \l_{ij2}\rvert^2 F(E_j,\tilde{\nu}_{ei}) -\lvert \l_{ij2}\rvert^2 F(\tilde{E}_j,\nu_{ei})\right],
\eea
where $F(a,b)$ is defined as
\be
F(a,b) = \int_0^1 dx \frac{x^2-x^3}{m_\mu^2 x^2 + (m_{a}^2 - m_\mu^2) x + m_b^2(1-x)}.
\ee
The existing measurements of the decay rates like $\tau \rightarrow \mu \gamma$, $\tau  \rightarrow 3l$, and $\tau\rightarrow \mu \nu \bar{\nu}$ can give constraints on $\l^{5}_{i2k}$ and $\l^{5}_{ij2}$ separately in combination with some other independent couplings \cite{Kim:2001se}. Assuming a hierarchy between $m_{\tilde{E},\tilde{e}}$ and $m_{\tilde{\nu}_{E},\tilde{\nu}_{e}}$, as a benchmark taking $m_{\tilde{E},\tilde{e}}\sim 700$ GeV and $m_{\tilde{\nu}_{E},\tilde{\nu}_{e}}\sim 250$ GeV,  we find that the current experimental data can be explained with less than order unity values of the couplings. Interestingly, in the presence of a mixing between left- and right-handed leptoquarks $(\tilde{h}_{L,R})$, it is possible to enhance the leptoquark contribution significantly to explain the data even without the slepton induced contributions.
%%%%%%%%%%%%%%%%%%%%%%%%%%%%%%%%%%%%%%%%%%%%%%%%%%%%%%%%%%%%%%%%%%%%%%%%%%%%%%%%%%%%%%%%%%%%%%%%%%%%%%%%%%%%%%%%%%%%%%%%%%%%%%%%%%%%%%%%%%%%%%%%%%%%%%%%%%%%%%%%%%%%%%%%%%%%%%%%%%%%%%%%

{\uppercase\expandafter{\romannumeral 6.\relax} \bf{Comments on lepton flavor violating (LFV) Higgs decay $h\to\tau^\pm\mu^\mp$ and constraints from $\tau\to \mu\gamma$ process---}}
In the SM, at tree level, LFV Higgs decays are absent and at loop level, they are highly suppressed by the GIM mechanism and small neutrino masses. Consequently, an observation of LFV Higgs decays with significant branching fractions would indicate NP contributions. Interestingly, the CMS Collaboration \cite{Khachatryan:2015kon} has reported a $2.4\sigma$ deviation in the measurement of LFV Higgs decay branching fraction $\rm{Br}(h\rightarrow \mu \tau)=0.084^{+0.39}_{-0.37}\%$, using the $19.7$ $\rm{fb^{-1}}$ data at a center of mass energy $\sqrt{s}=8\tev$, which is consistent with the ATLAS Collaboration $20.3$ $\rm{fb^{-1}}$ data with relatively large uncertainties \cite{Aad:2015gha}. The scalar leptoquark can contribute to the $h\to\tau\mu$ decay through the terms with coefficients $\lambda^6_{32j} \lambda^3_{33j}$. In the presence of a mixing between $\tilde{h}_{L,R}$, the dominant contribution to the $h\to\tau^\pm\mu^\mp$ decay rate is obtained when there is a helicity flip in the internal fermion lines 
\cite{Dorsner:2015mja,Altmannshofer:2015esa,Baek:2015mea,Herrero-Garcia:2016uab} and the contribution turns out to be proportional to $\sqrt{|\lambda^6_{32j} \lambda^3_{33j}|^2 + |\lambda^6_{33} \lambda^3_{33}|^2}$, which is constrained from $\tau\to\mu\gamma$ branching ratio. Using the constraints from $\tau\to\mu\gamma$, we find the $h\to\tau^\pm\mu^\mp$ branching fraction turns out to be orders of magnitude smaller than what is required to explain the CMS data. 
%%%%%%%%%%%%%%%%%%%%%%%%%%%%%%%%%%%%%%%%%%%%%%%%%%%%%%%%%%%%%%%%%%%%%%%%%%%%%%%%%%%%%%%%%%%%%%%%%%%%%%%%%%%%%%%%%%%%%%%%%%%%%%%%%%%%%%%%%%%%%%%%%%%%%%%%%%%%%%%%%%%%%%%%%%%%%%%%%%%%%%

{\uppercase\expandafter{\romannumeral 7. \relax}\bf{Conclusions---}}
We have presented a minimal framework of a left-right symmetric gauge theory naturally accommodating leptoquarks, which can provide a unified explanation of the $B$-decay anomalies in $R_{D^{(\ast)}}$ and $R_{K}$ together with the anomalous muon magnetic moment, while being consistent with the constraints from the current measurements of (semi)leptonic decays and $B_{s}^0-\bar{B}_{s}^0$, $D^0-\bar{D}^0$ mixings. In this model both $R_{D}$ and $R_{D^{\ast}}$ anomalies can be explained via the exchange of scalar leptoquarks at tree level, while the $R_K$ data can be explained simultaneously using one loop diagrams induced by leptoquarks. The anomalous muon magnetic moment can also be addressed in this model without utilizing a nonzero right-handed coupling of leptoquark.
%%%%%%%%%%%%%%%%%%%%%%%%%%%%%%%%%%%%%%%%%%%%%%%%%%%%%%%%%%%%%%%%%%%%%%%%%%%%%%%%%%%%%%%%%%%%%%%%%%%%%%%%%%%%%%%%%%%%%%%%%%%%%%%%%%%%%%%%%%%%%%%%%%%%%%%%%%%%%%%%%%%%%%%%%%%%%%%%%%%%%%

\begin{acknowledgements}
{\bf{Acknowledgments---}}DD would like to thank Ivan Ni\v{s}and\v{z}ic for many useful communications. CH would like to thank Utpal Sarkar for many helpful discussions. GK would like to acknowledge Matthias Neubert for a helpful correspondence. The authors would also like to thank Avelino Vicente for a very useful communication.
\end{acknowledgements}
%%%%%%%%%%%%%%%%%%%%%%%%%%%%%%%%%%%%%%%%%%%%%%%%%%%%%%%%%%%%%%%%%%%%%%%%%%%%%%%%%%%%%%%%%%%%%%%%%%%%%%%%%%%%%%%%%%%%%%%%%%%%%%%%%%%%%%%%%%%%%%%%%%%%%%%%%%%%%%%%%%%%%%%%%%%%%%%%%%%%
% % % % % % % % % % % % % % % % % % % % % % % % % % % % % % % % % % % % % % % % % % % % % % % % % % % % % % % % % % % % % % % % % % % % % % % % % % % % % % % % % % % % % % % % % % % % % % % % % % % % % % % % % % % % % % % % % % % % % % % % % % % % % % % % % % % % % % % % % % % % % % % % % % % % % % % % % % % % % % % % % % % % % % % % % % % % % % % % % % % % % % % % % % % % % % % % % % % % % % % % % % % % % % % % % % % % % % % % % % % % %

\end{document}